\documentclass[letterpaper,aps,prl,10pt,superscriptaddress,showpacs,floats,nofootinbib,twocolumn,pdftex]{revtex4-1}
\usepackage{cmap}
\usepackage[latin1]{inputenc}
\usepackage[T1]{fontenc}
\usepackage{subfigure}
\usepackage{amsmath,amssymb,mathbbol}
\usepackage{graphicx,color}
\usepackage{hyperref}

\definecolor{rltred}{rgb}{0.75,0,0}
\definecolor{rltgreen}{rgb}{0,0.5,0}
\definecolor{rltblue}{rgb}{0,0,0.75}
\hypersetup{colorlinks,%
  hypertexnames=true,%
  pdfauthor={J. Feist, S. Nagele, C. Ticknor, B. I. Schneider, L. A. Collins, J. Burgd\"orfer},%
  pdftitle={Attosecond two-photon interferometry for doubly excited states of helium},%
  pdfhighlight=/I,%
  breaklinks=true,%
  linkcolor=rltred,%
  citecolor=rltgreen,%
  linktocpage=true}

\begin{document}

\renewcommand{\equationautorefname}{Eq.}
\renewcommand{\figureautorefname}{Fig.}

\newcommand{\Int}{\int\limits}
\renewcommand{\Re}{\operatorname{Re}}

\newcommand{\etal}{\emph{et~al.\@} }
\newcommand{\etalb}{\emph{et~al.}}
\newcommand{\ie}{i.e., }
\newcommand{\Schro}{Schr\"o\-din\-ger }
\newcommand{\eg}{e.g.\@ }
\newcommand{\cf}{cf.\@ }

\newcommand{\expval}[1]{\langle#1\rangle}
\newcommand{\abs}[1]{\left|#1\right|}

\newcommand{\bra}[1]{\langle#1|}
\newcommand{\ket}[1]{|#1\rangle}
\newcommand{\braket}[2]{\langle#1|#2\rangle}
\newcommand{\cra}[1]{(#1|}
\newcommand{\craket}[2]{(#1|#2\rangle}
\newcommand{\DES}[5]{{}_{#1}(#2\,#3)^{#4}_{#5}}
\newcommand{\DESa}[1]{\ket{2s#1p^+}}
\newcommand{\DESb}[1]{\ket{2s#1p^-}}

\newcommand{\cvec}[1]{\mathbf{#1}}
\newcommand{\op}[1]{\mathrm{\hat{#1}}}
\newcommand{\eqcomma}{\,,}
\newcommand{\eqstop}{\,.}

\newcommand{\dd}{\mathrm{d}}
\newcommand{\DE}{\Delta E}

\newcommand{\ev}{\,\text{eV}}
\newcommand{\nm}{\,\text{nm}}
\newcommand{\He}{\text{He}}
\newcommand{\Hede}{\ensuremath{\text{He}^{**}}}
\newcommand{\Hep}{\ensuremath{\text{He}^{+}}}
\newcommand{\Hepp}{\ensuremath{\text{He}^{++}}}
\newcommand{\Wcm}{\,\text{W}/\text{cm}^2}
\newcommand{\as}{\,\text{as}}
\newcommand{\fs}{\,\text{fs}}

\newcommand{\Hecs}{\op{H}_\text{ECS}}
\newcommand{\tff}{{\tau}}

\newcommand{\cvkk}{{\cvec k_1\cvec k_2}}
\newcommand{\cvK}{{\cvec K}}
\newcommand{\bbK} {\widehat{\beta}_\cvK^{\phantom{0}}}

\newcommand{\level}[3]{{}^{#1}\!{#2}^{\textrm{#3}}}

\title{Attosecond two-photon interferometry for doubly excited states of helium}

\author{J.~Feist} 
\email{jfeist@cfa.harvard.edu}
\affiliation{ITAMP, Harvard-Smithsonian Center for Astrophysics, 
		     Cambridge, Massachusetts 02138, USA}

\author{S.~Nagele}
\email{stefan.nagele@tuwien.ac.at}
\affiliation{Institute for Theoretical Physics, Vienna University of Technology, 1040 Vienna, Austria, EU}

\author{C.~Ticknor}
\affiliation{Theoretical Division, 
             Los Alamos National Laboratory, Los Alamos, New Mexico 87545, USA}

\author{B.~I.~Schneider}
\affiliation{Office of Cyberinfrastructure/Physics Division, National Science Foundation,
			 Arlington, Virginia 22230, USA}

\author{L.~A.~Collins}
\affiliation{Theoretical Division, 
             Los Alamos National Laboratory, Los Alamos, New Mexico 87545, USA}

\author{J.~Burgd\"orfer}
\affiliation{Institute for Theoretical Physics, Vienna University of Technology, 1040 Vienna, Austria, EU}

\date{\today}

\begin{abstract}

We show that the correlation dynamics in coherently excited doubly excited resonances of helium can be followed in real time by two-photon interferometry.
This approach promises to map the evolution of the two-electron wave packet onto experimentally easily accessible non-coincident single electron spectra.
We analyze the interferometric signal in terms of a semi-analytical model which is validated by a numerical solution of the time-dependent two-electron \Schro equation in its full dimensionality.

\end{abstract}

\maketitle

Advances in optical technologies and laser sources in the past decade led to the production of extreme ultraviolet (XUV) light pulses as short as $80$ attoseconds (1 attosecond = $10^{-18}$ seconds) \cite{HenKieSpi2001,SanBenCal2006,GouSchHof2008}. Thus, the direct exploration of the electronic dynamics in atoms, molecules and solids in the time domain came into reach. This advance initiated a whole new field, \emph{attosecond physics}, and several pioneering experiments exploiting the novel technologies have already been performed (see \cite{Corkum07,KliVra2008,KraIva2009} and references therein).
Most measurement protocols either realized or proposed rely up to now on an interplay of an attosecond XUV pulse and a few-cycle IR pulse with durations $\tau_{\mathrm{IR}} \gtrsim 5$\,fs. Sub-fs time resolution is achieved through the exquisite sub-cycle control over carrier-envelope phase (CEP) stabilized IR pulses with uncertainties as small as $\Delta \varphi \approx 10^{-2} T_0$ where $T_0$ is the period of the IR oscillation. 
However, the direct analogue to femtosecond pump-probe spectroscopy in chemistry on the attosecond scale, \ie excitation of an electronic wavepacket by an attosecond pump pulse followed by an attosecond probe pulse to take snapshots of the ensuing electronic motion remains to be accomplished.
One obvious difficulty is that current attosecond XUV pulses based on high-harmonic generation (HHG) had, up to now, insufficient intensity to efficiently realize multi-photon pump-probe protocols.
Very recently, however, significant increases in HHG efficiency have been reported~\cite{FerCalLuc2010,PopCheArp2010}.
Therefore, attosecond XUV-XUV pump-probe experiments, which have been dubbed the ``holy grail'' of attosecond physics \cite{KliVra2008}, will likely be realized in the near future opening up a new stage of attosecond science.

This experimental perspective challenges theory to identify observables readily accessible in the experiment that map out non-trivial wavepacket dynamics of correlated electronic motion.
The paradigm system for correlated electron dynamics in real time are manifolds of coherently excited doubly excited states (\ie resonances) in helium. 
Pioneering experiments \cite{StrMor1986,YamMilKra1986} utilized collision excitation by charged particles as pump and the velocity of post-collisional energy shifts as probe.
From the velocity dependence of the angular differential autoionization spectra (``PCI effects'') the time evolution of collective two-electron variables such as the dipole, $\expval{\vec r_1 + \vec r_2}$ (or Runge-Lenz vectors $\expval{\vec a_1 + \vec a_2}$), or the vibronic motion of the interelectronic angle $\sim \expval{\vec a_1 \cdot \vec a_2}$ could be identified \cite{BurMor1988}.
For a XUV-XUV pump probe scenario a few theoretical proposals to guide attosecond-pulse experiments have been put forward. 
Hu and Collins \cite{HuCol2006} proposed to map out the wavepacket in coherently \emph{singly excited} helium created by the pump pulse.
They performed ab-initio calculations for the double ionization by the probe pulse as a function of delay time $\tau$ and showed that the total double ionization signal oscillation directly mirrors the radial breathing motion in the singly-excited state manifold.
This scenario requires, however, a two-color XUV-XUV pump-probe sequence.
Morishita \etal \cite{MorWatLin2007} showed, within lowest order perturbation theory, that the correlated motion of the two electrons in a wavepacket among the \emph{doubly excited} states (DES) of helium can be resolved by an XUV-XUV pump-probe scheme provided that the full six-dimensional two-electron momenta of the ejected electrons are resolved in a kinematically complete experiment.

\begin{figure}[tbp]
   \includegraphics[width=\linewidth]{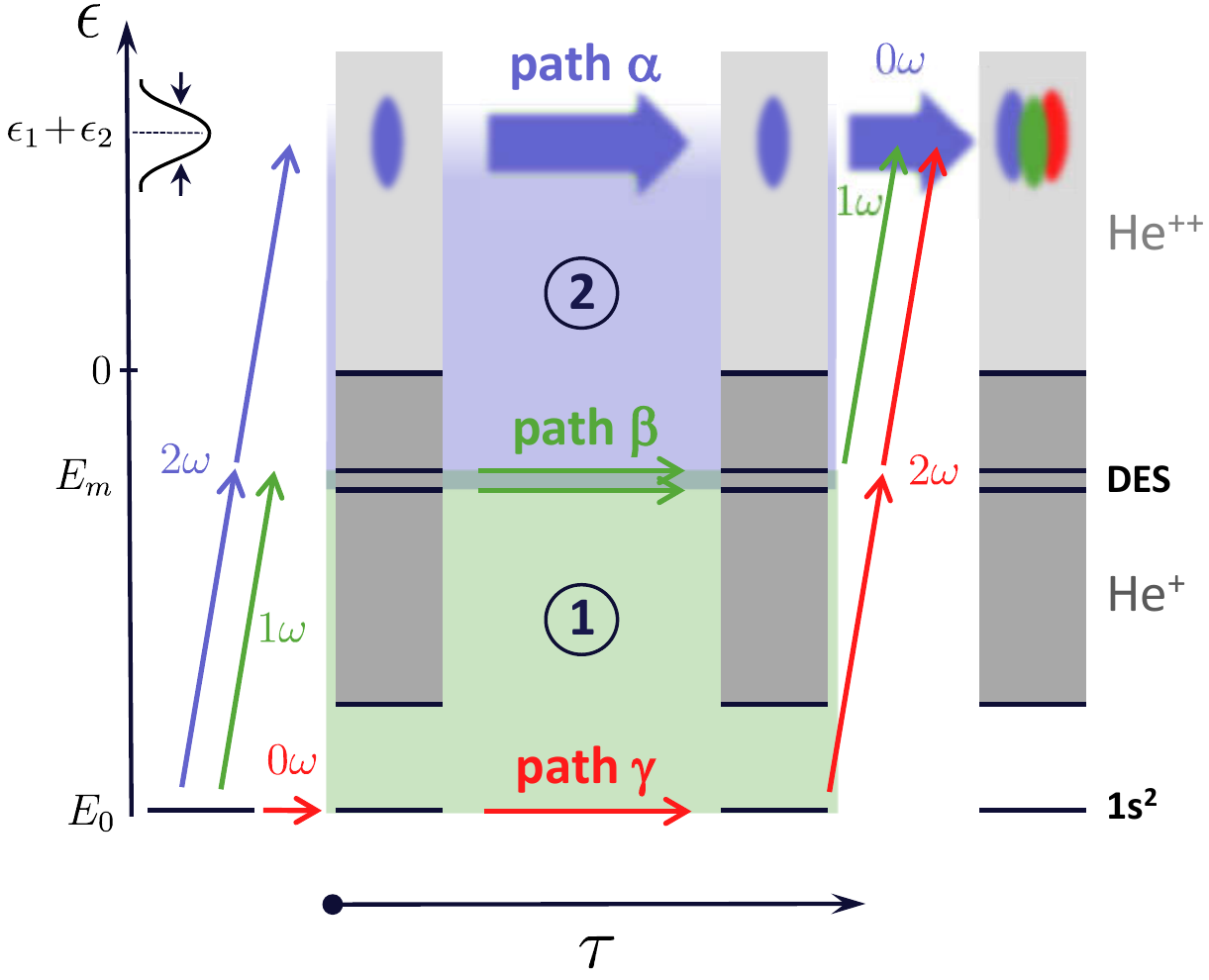}
\caption{Three-path interferometer for attosecond two-photon double ionization probing the coherent dynamics in doubly excited states (DES). The three paths $\alpha$, $\beta$, and $\gamma$ are represented by blue, green, and red arrows, respectively (see text). Interference areas $\DE \tau$: Area 1 (light-green) is delineated by (quasi-) bound states and is stable under average over $\epsilon_1$ (or $\epsilon_2$). Area 2 (in light-blue) is delineated by the energy $E=\epsilon_1+\epsilon_2$ of the two-electron continuum state and varies rapidly under variation of $\epsilon_1$ (or $\epsilon_2$).}
\label{fig:pp_levelscheme}
\end{figure}

In this letter we present a novel single-color XUV-XUV \emph{interferometric} pump-probe protocol that allows to follow the correlated two-electron motion in doubly excited states in real time by observing only (relatively) easily accessible integral and non-coincident experimental observables. 
To map out the electronic dynamics we exploit the interference between three two-photon double ionization pathways (see \autoref{fig:pp_levelscheme}) in a fashion which greatly enhances the observable signal.

We solve for the proposed scenario the time-dependent \Schro equation in its full dimensionality, including electronic correlations without further approximations (see \cite{PazFeiNag2011}).
The numerical parameters were chosen to ensure convergence.
The XUV pulses have a $\sin^2$ envelope with total duration of $1\fs$, a FWHM of intensity of $390\as$, and a central energy $\omega$ of $65.3\ev$.
All calculations presented in the following were performed for peak intensities of $10^{12}\Wcm$ for rapid numerical convergence.
For the experiment, values close to $10^{15}\Wcm$ would be desired.
We have explicitly checked that our results remain valid at such intensities, three-photon processes and ground state depletion are still negligible.

\begin{figure}[tbp]
    \centering
    \includegraphics[width=0.8\linewidth]{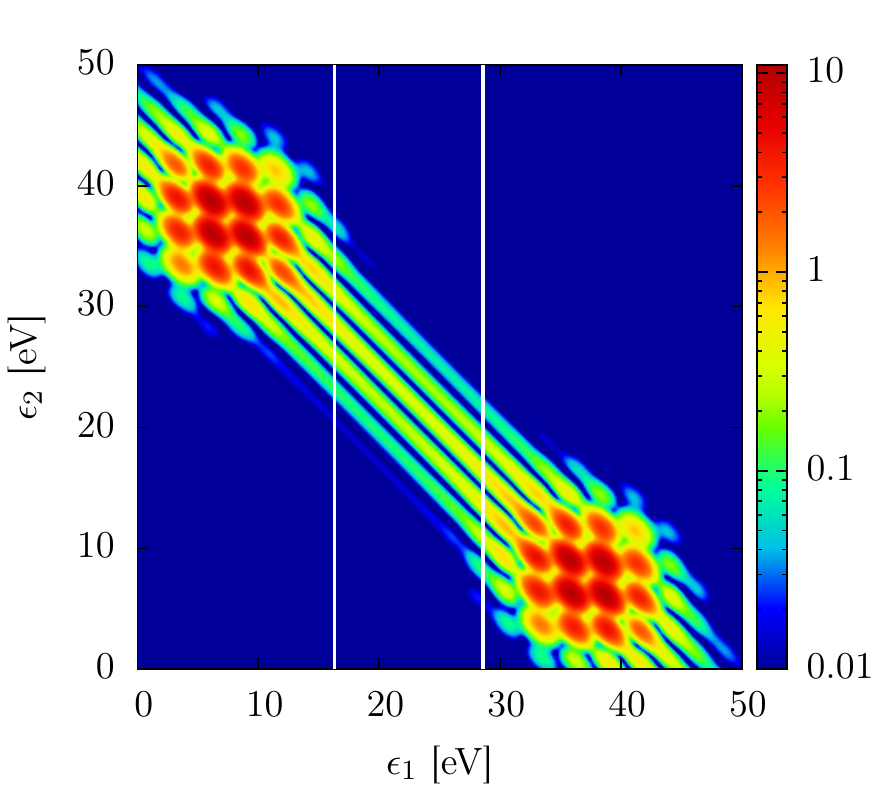}
\caption{Two-photon double ionization spectrum in the $(\epsilon_1,\epsilon_2)$ plane for a pump-probe sequence of two $1\fs$ $\sin^2$ $20\nm$ pulses 
with a peak-to-peak delay of $\tau=1500\as$. 
The white lines delimit the spectral window without contamination by sequential contributions. The diagonal oscillations 
in $\epsilon_1+\epsilon_2$ result from the interference between pathways $\alpha$ and $\beta+\gamma$. The complex interference pattern within the ``sequential'' peaks at $(\omega - I_1, \omega - I_2)$
is primarily due to interference between $\alpha$, $\gamma$, and the sequential pathway with one photon from each pulse (see text).
\label{fig:1s1s_1fs20nm_1fs20nm_2000as_wf_ph2}}
\end{figure}

The present attosecond two-photon pump-probe sequence (\autoref{fig:pp_levelscheme}) of DES can be viewed as a three-path interferometer, 
with the time delay $\tau$ between the pulses corresponding to the ``arm length'' of the interferometer.
Path $\alpha$ corresponds to two-photon double ionization by the pump pulse which has been the subject of a large number of recent investigations (see \eg \cite{PazFeiNag2011} and references therein). 
Path $\gamma$ is its replica induced by the probe pulse delayed by a time interval $\tau$ relative to the pump pulse. 
The intermediate path $\beta$ represents a proper pump-probe sequence where the first one-photon transition coherently excites a wavepacket of an ensemble of doubly excited states whose time evolution is then probed by double ionization by the second photon after the delay time $\tau$. 
Two specific features of this three-path interferometer, which displays a complex fringe pattern in the ($\epsilon_1,\epsilon_2$) plane of final energies of electron $1$ and $2$ (\autoref{fig:1s1s_1fs20nm_1fs20nm_2000as_wf_ph2}), are key to resolving the DES wavepacket dynamics.
First, path $\alpha$ represents a ``fuzzy'' slit. The interference phase $\DE \tau$ represented by the area enclosed in the $E-t$ diagram (\autoref{fig:pp_levelscheme}) between path $\alpha$ and any other path rapidly varies over the Fourier width of the total final energy, $\epsilon_1+\epsilon_2$, in the continuum (along the diagonal in \autoref{fig:1s1s_1fs20nm_1fs20nm_2000as_wf_ph2}).
Any partial trace over unobserved variables, \eg the energy of one electron, will wipe out any interference fringes associated with path $\alpha$ and will result in an incoherent and $\tau$-independent background contribution to the observed electron spectra.

\begin{figure}[tbp]
    \centering
    \includegraphics[width=\linewidth]{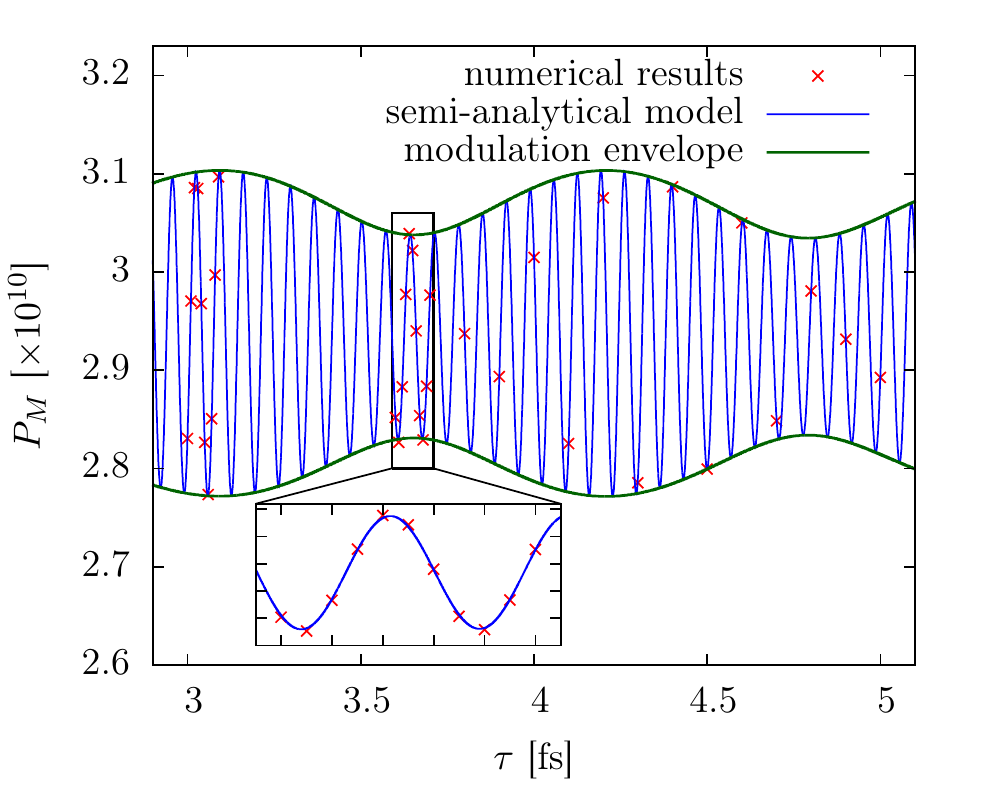}
\caption{Yield of restricted one-electron spectrum ($16.3\ev < \epsilon <28.6\ev$) integrated over all emission angles resulting from double ionization of He by a pump-probe sequence of a $20\nm$ pump -- $20\nm$ probe setup from the $1s^2$ singlet state as a function of delay time $\tau$ between pump and probe. Crosses: full numerical solution of the TDSE; blue line: semi-analytical model \autoref{eq:di_prob_model} including doubly excited resonances $\DESa{n}$, $n=2-5$, and $\DESb{3}$ as intermediate states; 
green line: envelope of the modulation of the fast oscillation between pathways $\beta$ and $\gamma$.
\label{fig:1s1s_1fs20nm_1fs20nm_e1_070_095_yield}}
\end{figure}  

For energies close to $(\epsilon_1,\epsilon_2) = (\omega - I_1, \omega - I_2)$ and its exchange symmetric partner $(\omega - I_2, \omega - I_1)$ where $I_1$ and $I_2$ are the ionization thresholds of $\He(1s^2)$ and $\Hep(1s)$, the additional pathway of sequential two-photon ionization, creating first $\Hep(1s)$ by the pump and then $\Hepp$ by the probe (omitted from \autoref{fig:pp_levelscheme} for clarity) gives rise to additional rapidly oscillating fringes within the Fourier broadened ``sequential peaks'' (see \autoref{fig:1s1s_1fs20nm_1fs20nm_2000as_wf_ph2}).
They can be removed by choosing an appropriate spectral window for the one-electron energies $(\omega - I_2) < \epsilon < (\omega - I_1)$ within the ``sequential'' peaks.
Focusing in the following on this energy window and integrating over the energy of the second electron leaves us with interference fringes that are exclusively determined by the phases, $\phi_m=(E_m^{\beta}-E_0^{\gamma})\tau$. The enclosed area (\autoref{fig:pp_levelscheme}) is delimited by the two sharp boundaries of the quasi-bound states of resonances (pathway $\beta$) and by the ground state $\He(1s^2)$ with energy $E_0$ (pathway $\gamma$).
Since the DES wavepacket encompasses several resonances with energies $E_m^{\beta}\,(m=1,\ldots)$,
the resulting interference fringes will display a fast oscillation on the attosecond scale given by the average phase $\expval{\phi_m(\tau)}$, and a slow modulation on a much longer time scale, $\phi_m(\tau)-\phi_{m'}(\tau)$, an example of which is shown in \autoref{fig:1s1s_1fs20nm_1fs20nm_e1_070_095_yield}.
This interference with the reference wave (pathway $\gamma$) may appear as an unwanted background signal that overshadows the pump-probe pathway $\beta$ but, instead, turns out to be the second key ingredient for improving the visibility of the coherent dynamics along pathway $\beta$.

The analysis of the interference signal is facilitated by a simple semi-analytical model extending a similar treatment for excited bound states to resonances \cite{MauRemSwo2010,ChoJiaMor2010}.
In this model for the two-photon interferometry we exploit the fact that only a limited number of states contributes to
the (differential) double ionization signal within the energy range of interest. 
These are the initial state (path $\gamma$) and the intermediate DES within the pump pulse bandwidth (path $\beta$). 
The latter are resonances embedded in the continuum of $\Hep$.
While the full numerical solution does not invoke an explicit representation of the DES, 
the model as well as the extraction of physical observables of the correlated dynamics, $\expval{O}_{\mathrm{DES}}$,
are facilitated by the explicit calculation of the DES. 
They are determined by an \emph{exterior complex scaling} (ECS) transformation of the Hamiltonian (\cf\cite{MccBaeRes2004} and references therein).
DES can not generally be described by independent-particle configurations, but require collective quantum numbers (\cf\cite{TanRicRos2000} and references therein).
In the current setup, only a restricted set of DES with $\level1Pe$ symmetry is accessed, and we use the traditional but imprecise labels $\ket{2snp^\pm}$ for brevity.
Correlated two-electron dynamics unfolds in the quasi-bound part of the resonances, the lifetime $\Gamma^{-1}$ of which typically exceeds $10\fs$.
To coherently excite a manifold of $\level1Pe$ doubly excited states in a one-photon transition from the $\He(1s^2)$ ground state,
photon energies $\omega_{\mathrm{XUV}} \gtrsim 60\ev$ (or wavelength $\lesssim 20\nm$) are required.
The spectral width should be of the order of a few $\ev$ corresponding to an attosecond pulse with $t_{\mathrm{XUV}} \lesssim 1 \fs$.
Assuming that pump and probe pulses are temporally separated,
the final wavepacket can be written as
\begin{gather}\label{eq:psi_final}
\ket{\psi_f} = \op{U}^{(2)} e^{-i \Hecs \tff}\, \op{U}^{(1)} \ket{\psi_0} \eqcomma
\end{gather}
where $\Hecs$ is the field-free ECS Hamiltonian, $\op{U}^{(i)}$ is the time evolution
operator associated with the $i$th pulse (1=pump, 2=probe), and
$\tff$ is the duration of the field-free evolution between the pulses.
We spectrally decompose the field-free propagation operator $e^{-i\Hecs \tff}$ and retain only 
the relevant intermediate states, the initial state $\ket{\gamma}\equiv\ket{\psi_0}$ (pathway $\gamma$) 
and intermediate DES $\ket{\beta^m}\equiv\ket{2snp^\pm}$ (pathway $\beta$). Up to a global phase, 
the double ionization amplitude at the conclusion of the probe pulse is
\begin{equation}\label{eq:di_prob_model}
\braket\cvK{\psi_f} = \gamma_\cvK + \sum_{m} e^{-i \DE_m \tff} \beta_\cvK^m \eqcomma
\end{equation}
with $\cvK\equiv(\cvkk)$, $\DE_m = E_m - E_0$, and $g_\cvK = \bra\cvK \op{U}^{(2)} \ket{g}\cra{g} \op{U}^{(1)} \ket{\psi_0}$ (for $g=\beta,\gamma$).
 In the intermediate step, we use the modified inner product $\craket{n}{m}=\braket{n^*}{m}$, as $\Hecs$ is 
complex symmetric. 
$E_m$ (and thus $\DE_m$) is \emph{complex}, with the imaginary part describing the decay of the DES. 

\begin{figure*}[tb]
  \includegraphics[width=0.63\linewidth]{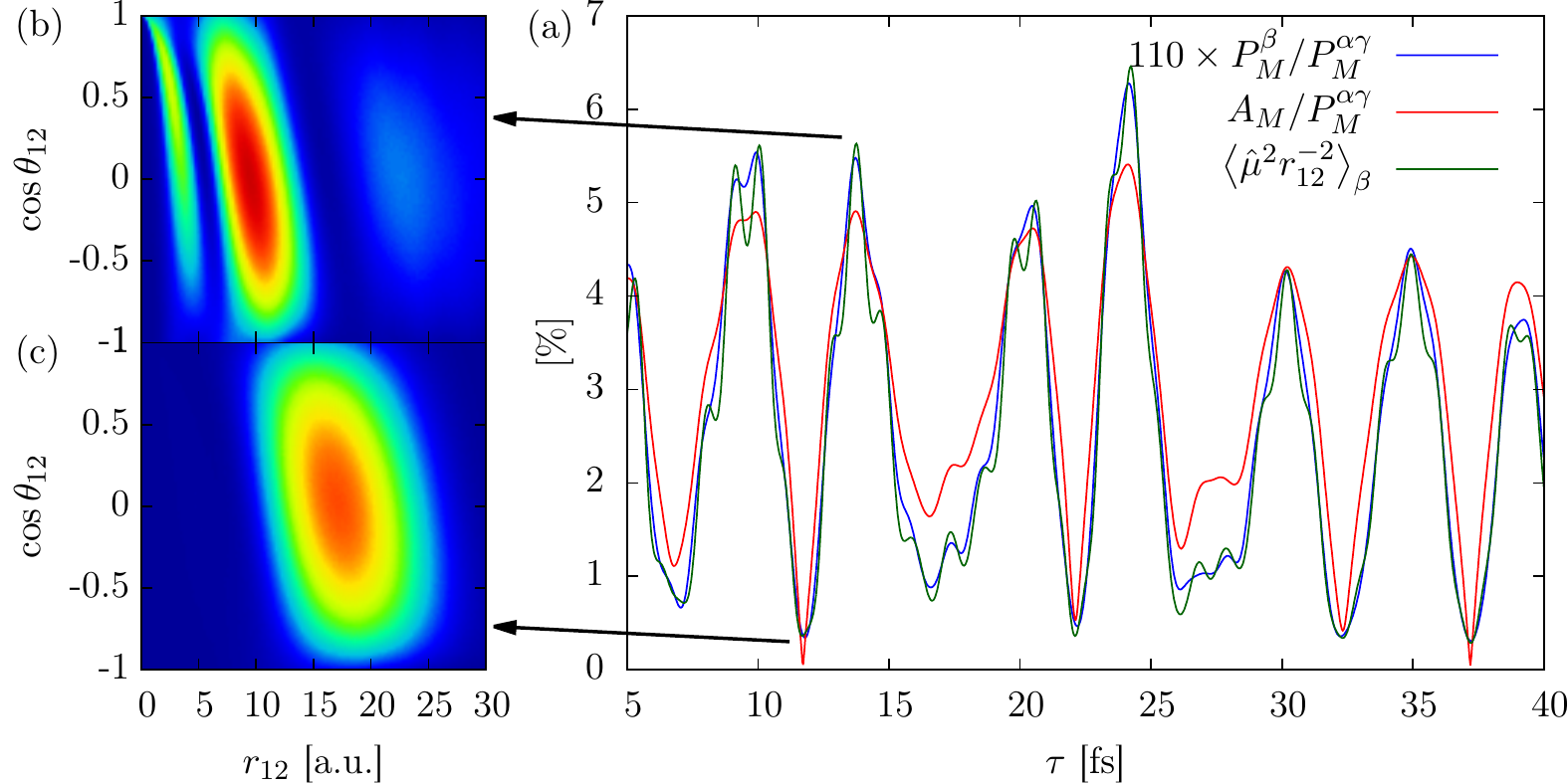}\hfill
  \begin{minipage}[b]{0.35\linewidth}
  \caption{$(a)$ Yield $P_M^\beta$ from DES and modulation $A_M$, shown as ratios to the background yield $P_M^{\alpha\gamma}=2\int_M |\gamma_\cvK|^2 \dd\cvK$ from paths $\alpha$ and $\gamma$, for the restricted one-electron spectrum ($17.7\ev\!<\!\epsilon\!<\!30.0\ev$) from double ionization integrated over all emission angles, compared with the DES expectation value $\expval{\op \mu^2 r_{12}^{-2}}_\beta$. The pulses ($\sin^2$ shape with $2\fs$ total duration, central wavelength $19\nm$) coherently excite $\DESa{n}$ ($n=3-8$) with appreciable probability.
Two-dimensional projections of the two-electron wavepacket on the $(r_{12},\cos\theta_{12})$ plane at the maximum $(b)$ and minimum $(c)$ of the modulation $A_M$.}
\label{fig:1s1s_2fs19nm_and_2fs19nm_yields}
  \end{minipage}
\end{figure*}

Each of the mixed terms $\sim$\,$\gamma_\cvK^* \beta_\cvK^m$ in the probability $P_\cvK=\abs{\braket\cvK{\psi_f}}^2$ oscillates with frequencies $\Re(\DE_m)$ corresponding to periods of $\approx$\,$70\as$. 
The superposition of several terms, $\bbK(\tff)=\sum_m e^{-i \DE_m \tff} \beta_\cvK^m$, to which only resonances within the bandwidth of the pump pulse contribute, leads to a modulation with frequencies $\Re(E_m - E_{m'})$ corresponding to periods on the (multi-)femtosecond scale (\autoref{fig:1s1s_1fs20nm_1fs20nm_e1_070_095_yield}) given by the energy spacing between the DES.
Since $|\bbK|^2$ is proportional to the product of XUV one-photon double excitation probabilities and the double ionization probability from the weakly bound doubly excited states, it is three to four orders of magnitude smaller than the two-photon double ionization from the ground state $\sim\!|\gamma_\cvK|^2$. Consequently, the interferometric signal $\sim\!\Re(\gamma_\cvK^* \bbK)$ is enhanced by orders of magnitude compared to the true pump-probe signal $|\bbK|^2$. 
It appears as the modulation amplitude relative to an approximately constant background $\sim\!2|\gamma_\cvK|^2$ 
(where the factor two takes the contribution $|\alpha_\cvK|^2\approx|\gamma_\cvK|^2$ into account).
The latter can be independently determined from the measurement of the pump signal alone in the absence of a probe pulse.
In turn, the modulation amplitude $A_M$ follows from
\begin{equation}\label{eq:di_integrated_prob}
A_M(\tau) = 4\abs{ \int_M \gamma_\cvK^* \bbK \dd\cvK }
\end{equation}
where $M$ is the region of final-state electron momenta integrated over:
an energy window for electron 1 (\autoref{fig:1s1s_1fs20nm_1fs20nm_2000as_wf_ph2}), all emission angles of electron 1 and all vectorial momenta of electron 2. 
The modulation $A_M(\tau)$ is the experimentally accessible signal monitoring the wavepacket dynamics in the collectively excited DES manifold, 
and agrees remarkably well with the (experimentally inaccessible) direct contribution from the DES 
pump-probe path $\beta$, $P_M^\beta = \int_M |\bbK|^2 \dd\cvK$ (\autoref{fig:1s1s_2fs19nm_and_2fs19nm_yields}).
This good agreement results from the fact that two-photon double ionization in a single pulse produces a ``well-behaved'' reference wave.

It is now of crucial importance to identify the expectation values of observables within the DES manifold with which the probe signal $A_M$ approximately correlates.
Clearly, because of the dipole selection rules the two-photon XUV pump-probe scenario will give access to observables differently from those monitored by charged-particle collisions \cite{StrMor1986,YamMilKra1986,BurMor1988}. 
Key is the observation that double ionization of DES by absorption of a single photon from the probe pulse is mediated by final state correlation.
To lowest order perturbation theory, this is the well-known two-step-one (TS1) process frequently invoked for both photoionization and charged particle ionization \cite{CarKel1981,VegBur1990}.
Accordingly, one electron absorbs the photon energy and ejects the second electron by a collisional Coulomb interaction in the exit channel.
The amplitude of this process is proportional to $\bra{\cvK_{(0)}}r_{12}^{-1} \op\mu\ket{\psi_\beta}$,
where $\cvK_{(0)}$ represents the uncorrelated final two-electron continuum state, $\op\mu = p_{z,1} + p_{z,2}$ is the dipole transition operator 
and $\ket{\psi_\beta}$ is the DES part of the intermediate wave packet.
The probability for one-photon double ionization of DES with final momenta in the restricted region is therefore
\begin{equation}\label{eq:P_M}
P_M^\beta(\tau) \propto \Int_M  \dd\cvK \bra{\psi_\beta}\op\mu r_{12}^{-1}\ket{\cvK_{(0)}}\bra{\cvK_{(0)}}r_{12}^{-1} \op\mu\ket{\psi_\beta} \eqstop
\end{equation}
Invoking the closure approximation $\int_M\!\ket{\cvK_0}\bra{\cvK_0}\dd\cvK\!\approx\!\mathbb{1}$, \autoref{eq:P_M} reduces to the expectation value
\begin{equation}\label{eq:P_M_approx}
P_M^\beta(\tau) \propto \bra{\psi_\beta} \op\mu^2 r_{12}^{-2} \ket{\psi_\beta},
\end{equation}
\ie the dipole-weighted square of the electron-electron interaction.
\autoref{eq:P_M_approx} agrees remarkably and, in view of the a priori poorly justified closure approximation, surprisingly well with the simulated modulation signal $A_M$ (\autoref{fig:1s1s_2fs19nm_and_2fs19nm_yields}) for the complex modulation pattern resulting from a pump-probe sequence with a central wavelength of $19\nm$.
We note that leaving out the dipole operators, \ie using the expectation value $\bra{\psi_\beta} r_{12}^{-2} \ket{\psi_\beta}$ works equally well. \autoref{fig:1s1s_2fs19nm_and_2fs19nm_yields} clearly represents signatures of the time-resolved correlation dynamics appearing in the non-coincident single-electron spectrum.

The numerical simulation allows to explore the correlated two-electron dynamics, the projection of which onto the single-electron spectrum is monitored by $A_M$.
Snapshots in the $(r_{12}, \cos\theta_{12})$ plane reveal that maxima (minima) in $A_M$ are associated with minima (maxima) in the inter-electronic  separation rather than with the one-electron distance from the nucleus. The latter would be the hallmark of mean-field (or independent particle) processes.

In summary, we have shown how correlated dynamics in doubly excited states of helium
can be accessed by two-photon interferometry with identical attosecond pulses. 
Supported by a full numerical solution of the \Schro equation, 
we have shown that contributions from two-photon absorption within a single pulse 
provide a \emph{reference wave}
that the signal of interest interferes with and that greatly enhances the observable signal.
The present protocol may provide an avenue for directly observing correlation dynamics with attosecond pulses available presently or in the near future without coincidence requirements.

The authors wish to thank Renate Pazourek for valuable discussions. We acknowledge support by the FWF-Austria, grants No.\ SFB016 and P21141-N16 (S.N. \& J.B.) 
and by the NSF through a grant to ITAMP (J.F.). C.T.\ and L.A.C.\ acknowledge support from LANL, which is operated by LANS,
LLC for the NNSA of the U.S. DOE under Contract No. DE-AC52-06NA25396.
The computational results have been achieved using the Vienna Scientific Cluster, Institutional Computing resources at 
Los Alamos National Laboratory, and NSF TeraGrid resources provided by NICS and TACC under grant TG-PHY090031.

\end{document}